\documentclass[sigconf]{acmart}

\usepackage{amsmath}
\usepackage{amsthm}
\usepackage{enumitem}
\usepackage{color}
\usepackage{makecell}
\usepackage{subcaption}
\usepackage{graphicx}
\usepackage{xcolor}
\usepackage{listings}
\usepackage[utf8]{inputenc}

\definecolor{deepblue}{rgb}{0,0,0.5}
\definecolor{deepred}{rgb}{0.6,0,0}
\definecolor{deepgreen}{rgb}{0,0.7,0}
\definecolor{codepurple}{rgb}{0.58,0,0.82}

\DeclareFixedFont{\ttb}{T1}{txtt}{bx}{n}{9} 
\DeclareFixedFont{\ttm}{T1}{txtt}{m}{n}{9}  

\lstdefinestyle{java-code-style}{
language=Java,
basicstyle=\footnotesize\tt,
numberstyle=\tiny\color{deepblue},
morekeywords={self},              
keywordstyle=\color{deepred},
emph={MyClass,__init__},          
emphstyle=\ttb\color{deepred},    
stringstyle=\color{deepgreen},
frame=tb,                         
commentstyle=\selectfont\color{deepgreen}, 
showstringspaces=false
}

\lstdefinestyle{python-style-for-table}{
language=Python,
basicstyle=\tiny\tt,
numberstyle=\tiny\color{deepblue},
morekeywords={self},              
keywordstyle=\color{deepred},
emph={MyClass,__init__},          
emphstyle=\ttb\color{deepred},    
stringstyle=\color{deepgreen},
frame=none,                         
showstringspaces=false,
otherkeywords={1, 2, 3, 4, 5, 6, 7, 8, 9, 0},
breaklines=true,
commentstyle = \color{olive},
literate=%
  {+}{{{\tiny\color{deepblue}+}}}1
  {-}{{{\tiny\color{deepblue}-}}}1
  {*}{{{\tiny\color{deepblue}*}}}1
  {/}{{{\tiny\color{deepblue}/}}}1
  {=}{{{\tiny\color{deepblue}=}}}1
  {:}{{{\tiny\color{deepblue}:}}}1
}

\usepackage{multirow}

\usepackage[linesnumbered, figure, vlined]{algorithm2e}

\definecolor{dkgreen}{rgb}{0,0.46,0.25}
\definecolor{gray}{rgb}{0.5,0.5,0.5}
\definecolor{mauve}{rgb}{0.58,0,0.52}

\SetCommentSty{mycommfont}
\newcommand{\afgnn}{AFGNN}
\newcommand{\mugnn}{AFGNN}
\usepackage{url}

\newcommand{\mugcn}{AFGCN}
\newcommand{\murgcn}{AFRGCN}

\newcommand{\tbox}[1]{{\setlength{\fboxrule}{1.5pt}\fcolorbox{red}{white}{\texttt{#1}}}}

\newsavebox{\tempbox}

\AtBeginDocument{%
  }

\setcopyright{acmlicensed}
\copyrightyear{2018}
\acmYear{2018}
\acmDOI{XXXXXXX.XXXXXXX}
\acmConference[MSR'26]{Make sure to enter the correct
  conference title from your rights confirmation email}{April 13--14,
  2026}{Rio de Janeiro, Brazil}
\acmISBN{978-1-4503-XXXX-X/2018/06}

\copyrightyear{2026}
\acmYear{2026}
\setcopyright{cc}
\setcctype{by}
\acmConference[MSR '26]{23rd International Conference on Mining Software Repositories}{April 13--14, 2026}{Rio de Janeiro, Brazil}
\acmBooktitle{23rd International Conference on Mining Software Repositories (MSR '26), April 13--14, 2026, Rio de Janeiro, Brazil}
\acmDOI{10.1145/3793302.3793372}
\acmISBN{979-8-4007-2474-9/2026/04}

\begin{document}

\title{\mugnn{}: API Misuse Detection using Graph Neural Networks
and Clustering}

\author{Ponnampalam Pirapuraj}
\affiliation{%
  \institution{IIT Hyderabad}
  \city{Hyderabad}
  \state{Telangana}
  \country{India}
}
\email{cs23resch16001@iith.ac.in}
\orcid{0000-0003-0056-7683}

\author{Tamal Mondal}

\affiliation{%
  \institution{Oracle}
  \city{Hyderabad}
  \state{Telangana}
  \country{India}
}
\email{tamalmondal495@gmail.com}
\orcid{0009-0008-7901-9877}

\author{Sharanya Shathavelli}
\affiliation{%
  \institution{Yokogawa Electric}
  \city{Tokyo}
  \state{Tokyo}
  \country{Japan}
}
\email{sharanyagupta@alumni.iith.ac.in}
\orcid{0009-0002-3075-5597}

\author{Akash Lal}
\affiliation{%
  \institution{Microsoft Research}
  \city{Bangalore}
  \state{Karnataka}
  \country{India}
}
\email{akashl@microsoft.com}
\orcid{0009-0002-4359-9378}

\author{Somak Aditya}
\affiliation{%
  \institution{IIT Kharagpur}
  \city{Kharagpur}
  \state{West Bengal}
  \country{India}
}
\email{aditya.somak@gmail.com}
\orcid{0000-0002-0113-2545}

\author{Jyothi Vedurada}
\affiliation{%
  \institution{IIT Hyderabad}
  \city{Hyderabad}
  \state{Telangana}
  \country{India}
}
\email{jyothiv@cse.iith.ac.in}
\orcid{0000-0002-5911-6011}







\renewcommand{\shortauthors}{Pirapuraj et al.}

\begin{abstract}
  Application Programming Interfaces (APIs) are crucial to software development, enabling integration of existing systems with new applications by reusing tried and tested code, saving development time and increasing software safety. 
In particular, the Java standard library APIs, along with numerous third-party APIs, are extensively utilized in the development of enterprise application software.
However, their misuse remains a significant source of bugs and vulnerabilities.
Furthermore, due to the limited examples in the official API documentation, developers often rely on online portals and generative AI models to learn unfamiliar APIs, but using such examples may introduce unintentional errors in the software.
In this paper, we present \mugnn{}, a novel Graph Neural Network (GNN)-based framework for efficiently detecting API misuses in Java code. 
\mugnn{} uses a novel API Flow Graph (AFG) representation that captures the API execution sequence, data, and control flow information present in the code to model the API usage patterns.  
\mugnn{} uses self-supervised pre-training with AFG representation to effectively compute the embeddings for unknown API usage examples and cluster them to identify different usage patterns. Experiments on popular API usage datasets show that \mugnn{} significantly outperforms state-of-the-art small language models and API misuse detectors.

\end{abstract}




\maketitle

\section{Introduction}
Developers are often unsure how to use certain library APIs (Application Programming Interfaces), even when documentation is available~\cite{API-usage-difficulty}.
To find API usage examples, they commonly use online portals~\cite{stackoverflow} or generative AI tools~\cite{achiam2023gpt}. 
Using these examples may introduce unintentional errors and vulnerabilities to the software due to the prevalence and severity of API misuse in them~\cite{zhang2018code, zhong2024can}. 
For example, the code sample in Figure~\ref{fig:br-readLine-usage}, retrieved from the commit history of a GitHub project~\cite{github-commit-link}, demonstrates the use of the {\tt BufferedReader.readLine()} API method at Line~\ref{ln:API_call_br}.
A correct usage is shown where the {\tt BufferedReader} object {\tt br} is properly closed using {\tt br.close()} at Line~\ref{ln:API_call_to_close}, ensuring proper resource management.
In contrast, the misuse omits this close operation, potentially leading to a resource leak vulnerability.
These actual examples of correct and incorrect {\tt BufferedReader} API usage from GitHub commits highlight the occurrence of API misuse in real-world code and the importance of ensuring API usage correctness.

\lstset{style=java-code-style}


\begin{figure}[t]
\centering
\begin{subfigure}{\linewidth}
\begin{lstlisting}[style=java-code-style, numbers=left, numberstyle=\tiny, stepnumber=1, xleftmargin=2em, numbersep=3pt, escapechar=|, breaklines=true, breakatwhitespace=true, basicstyle=\ttfamily\footnotesize]
public static void main(String[] args) throws Exception {
    BufferedReader br = new BufferedReader(new InputStreamReader(System.in));
    StringTokenizer st = new StringTokenizer(br.readLine()); |\label{ln:API_call_br}|
    StringBuilder sb = new StringBuilder();
    ...
    for (int i = 1; i <= N; i++) {
        sb.append(count[i]).append(" ");
    } 
    System.out.println(sb.toString());
    /* BufferedReader is not closed */
}
\end{lstlisting}
\caption{API misuse example of {\tt BufferedReader.readLine()}}
\label{fig:misuse-example}
\end{subfigure}
\begin{subfigure}{\linewidth}
\begin{lstlisting}[language=Java, numbers=left, numberstyle=\tiny, xleftmargin=2em, stepnumber=1, numbersep=3pt, escapechar=|, breaklines=true, breakatwhitespace=true, basicstyle=\ttfamily\footnotesize]
 public static void main(String[] args) ... {
    BufferedReader br = new BufferedReader(new InputStreamReader(System.in));
    StringTokenizer st = new StringTokenizer(br.readLine());
    StringBuilder sb = new StringBuilder();
    ... //for loop not shown
    System.out.println(sb.toString());
    |\tbox{br.close();}||\label{ln:API_call_to_close}|
}
\end{lstlisting}
\caption{Correct usage example of {\tt BufferedReader.readLine()}}
\label{fig:correct-usage-example}
\end{subfigure}
\caption{Usage of {\tt BufferedReader.readLine()} API.}
\label{fig:br-readLine-usage}
\Description{}
\end{figure}

To this end, recent efforts have proposed tools and techniques~\cite{code-kernel, 9438601, lyu2021embedding, wen2019exposing} that can analyze and detect potential API misuse or suggest correct API usage examples.
These tools help minimize errors, improve security, and reduce development time, but they are limited in their ability to handle long code segments, to be computationally efficient, and to maintain accuracy.
For example, CodeKernel~\cite{code-kernel}, a graph kernel based approach, works well with API usage programs that are relatively short, but its accuracy drops significantly for longer code segments.

It is also possible to classify API misuse using very large language models (LLMs) such as GPT-4~\cite{achiam2023gpt}, and DeepSeek~\cite{bi2024deepseek} which offer strong reasoning capabilities for code understanding, but their high computational cost, hard to fine-tune nature limit their practical applicability in API-misuse detection. 
This motivates the development of lightweight, structure-aware approaches, such as small language models (LMs)~\footnote{Here, by ``{small} LMs'', we refer to smaller models (Million-level parameters, e.g. $125$M) like  GraphCodeBERT~\cite{guo2020graphcodebert} and UnixCoder~\cite{Unixcoder}, while by ``LLMs'', we refer to models (Billion-level parameters, e.g. $175$B) like  GPT~\cite{achiam2023gpt}.} and graph-based models, that can efficiently learn API-usage patterns without relying on massive compute or costly prompting.
Embeddings generated by small LMs can be used to detect API misuse.
However, small LMs such as CodeT5+~\cite{CodeT5+}, CodeBERT~\cite{feng-etal-2020-codebert}, GraphCodeBERT~\cite{guo2020graphcodebert} and UniXcoder~\cite{Unixcoder} even though good at providing task-agnostic embeddings, perform poorly in detecting API misuse because: (1) they do not take into account API-specific data and control-flow relationships but instead focus on the sequence of code tokens, (2) they suffer information loss for longer code snippets due to input length limitation.
Further, Graph Neural Network (GNN) based models like GraphCode2Vec~\cite{ma2022graphcode2vec} 
require expensive static analysis on the intermediate representation of the code to obtain the embeddings. 



To address these challenges, we propose \mugnn{}, a lightweight GNN-based framework that detects potential API misuse in input code efficiently.
\mugnn{} generates embeddings to capture API usage and then clusters the embeddings.
To model the API usage patterns, \mugnn{} uses a novel API Flow Graph (AFG) representation that captures the data flow, control flow, and API call sequence in the code that flows through the API call site. 
\mugnn{} uses self-supervised pre-training with AFG representation to effectively compute the embeddings for unknown API usage examples.
These embeddings are then clustered based on their similarity to identify the usage patterns.
A larger cluster indicates frequent API usage, suggesting correct use, while a smaller cluster denotes infrequent usage, indicating potential misuse.
Prior work~\cite{monperrus2013detecting, code-kernel, lindig2015mining, li2005pr, kang2021active} has shown that frequent patterns in real-world projects often correspond to correct API usage. 

Our pre-trained GNN model is a fraction of existing {small} LMs in terms of model size and so needs less memory and computation power. 
Further, our evaluation shows that \mugnn{} has superior performance in understanding API usage and detecting API misuse compared to state-of-the-art {small} LMs and misuse detectors.

The key contributions of this work are as follows:
\begin{itemize}
  \item A manually labelled API usage dataset with widely used Java APIs suitable for clustering API usage patterns.
  \item A new API Flow Graph (AFG) representation that captures the data flow, control flow, and API call sequences in a code snippet.
  \item \mugnn{}, a GNN-based model that uses AFG representation and self-supervised pre-training to learn flow-aware code embeddings, and clusters them to discover API usage patterns. 
  \item Extensive evaluation showing the effectiveness of our approach over state-of-the-art {small} LMs in API-usage clustering and misuse detection, along with comparison to existing misuse detectors on real-world examples.
 Ablation study assessing the impact of different \mugnn{} components and validating our design choices.
\end{itemize}

\section{Related Work and Background}
\label{s:back}

\begin{figure*}[htp]
\centering
\includegraphics[width=0.8\textwidth]{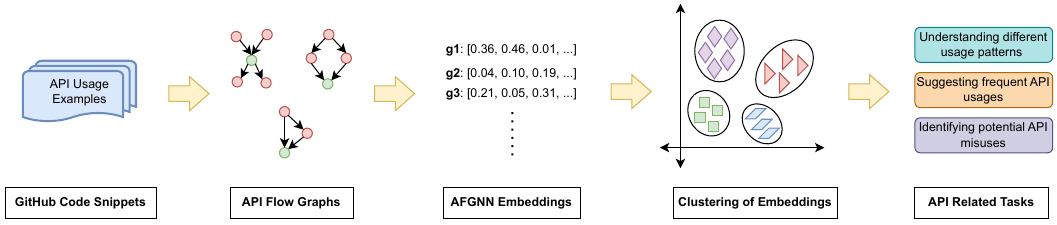}
\caption{Overview of \mugnn{}. Given method-level code snippets, we construct API Flow Graphs (AFGs), generate their embeddings using AFGNN, and cluster the embeddings to identify frequent usage patterns and potential misuses.}
\label{fig:mugnn-pipeline}
\Description{}
\end{figure*}

\noindent{\it Language Models and  GNNs for Code Understanding.}
Pre-trained language models such as  CodeBERT~\cite{feng-etal-2020-codebert}, GraphCodeBERT~\cite{guo2020graphcodebert}, UniXCoder~\cite{Unixcoder}, and CodeT5+~\cite{CodeT5+} are relatively small in size (million-parameter scale) and have shown strong performance in code understanding tasks. 
These encoder or encoder-decoder models process code syntax and semantics to generate embeddings that can be used to detect API misuse but they struggle with longer code snippets due to input length constraints.
GraphCodeBERT enhances embeddings via data-flow graphs, while UniXCoder integrates  abstract syntax trees and comments. 
However, the absence of control-flow modeling in GraphCodeBERT limits its effectiveness in capturing complex API usage patterns.
CodeT5+ leverages the T5 architecture with advanced objectives for good
performance.

Since program structures are inherently graph-based, with representations like Abstract Syntax Tree (AST), Control-Flow Graph (CFG), Data-Flow Graph (DFG)~\cite{davis1982data}, and Program Dependence Graph (PDG)~\cite{ferrante1987program},
Graph Neural Networks are well-suited for program analysis, having demonstrated remarkable success in learning representations that capture both local and global graph structures.
Different types of GNNs include Graph Convolutional Networks (GCNs)~\cite{kipf2016semi}, Graph Attention Networks (GATs)~\cite{velivckovic2017graph}, Relational Graph Convolutional Networks (RGCNs)~\cite{thanapalasingam2022relational}, and Relational Graph Attention Networks (RGATs)~\cite{busbridge2019relational}. 
GCNs extend the concept of convolution from grids to graphs, aggregating information from neighbouring nodes. 
GATs enhance this by applying attention mechanisms to weigh the importance of neighbouring nodes differently. 
RGCNs handle multi-relational data by incorporating edge types into the aggregation process, while RGATs combine relational information with attention mechanisms. 
GNNs iteratively update node representations through message passing, aggregating information from neighbour nodes, combining it with their own features, and updating their state. 
Key strategies in GNNs include the choice of aggregation functions, which can be mean, sum, or max operations, and the use of normalization techniques to stabilize training. 
A broader discussion on adapting GNNs for code representation can be found in the study by Allamanis et al.~\cite{allamanis2022graph}.  
As we represent input code as a graph (AFG), \mugnn{} uses GNN architecture to leverage GNNs' efficiency in understanding such graph structures.

Luo et al.~\cite{luo2022compact} use Compressed Abstract Graph (CAG), a  compact representation of AST, to represent code for vulnerability detection that preserves key structure and semantics while accelerating model training. 
They used MPNet~\cite{song2020mpnet} to get node embeddings, followed by a two-layer GNN with soft attention pooling. 
While they emphasize compressed graphs for vulnerability detection, \mugnn{} introduces AFGs by modeling data and control dependencies and API sequences for API Misuse Detection.
Zhou et al. ~\cite{wang2024suitable} extract task-specific subgraphs from Code Property Graphs (CPGs)~\cite{yamaguchi2014modeling} that integrate ASTs, control flow graphs, and program dependency graphs to capture comprehensive code semantics.
Nodes are encoded via operation types, function types, and semantic vectors from Word2Vec~\cite{mikolov2013efficient}, processed by a GCN with attention.
In contrast, \mugnn{} introduces API sequence modeling in AFGs.

\noindent{\it Graph-based Approaches for API Misuse Detection.}
Graph-based methods like Code-Kernel~\cite{code-kernel} and GraphCode2Vec~\cite{ma2022graphcode2vec} represent code as graphs for embedding.
Code-Kernel uses graph kernels but lacks control flow modeling and machine learning support (heuristic-based), limiting its applicability for detecting large, complex API usage patterns.
\mbox{GraphCode2Vec} fuses lexical and dependency embeddings using static analysis. 
While effective, it faces scalability challenges on large or complex codebases due to its costly static analysis. It also cannot be applied directly to our datasets, as it requires compilable Java code for inference, whereas our method-level data often contains partial, non-compilable code.
ADG-Seq2Seq~\cite{lyu2021embedding} uses API dependency graphs but omits control dependencies. 
In contrast, along with data dependencies, \mugnn{} models control flow and the API call sequences in AFGs.

Among rule-based and knowledge-driven methods, Xia Li et al.~\cite{9438601} conducted a large-scale study on GitHub bug-fix commits to detect specific API misuse types like guard conditions and exception handling, using fine-grained AST differencing and a lightweight intra-procedural analysis. 
However, their approach relies on manually crafted detection rules. 
MuDetect~\cite{MuDetect}, represents code with API-Usage Graphs (AUGs) and applies a greedy, semantic-aware subgraph mining and specialized graph matching, along with a ranking strategy, to improve precision and recall over prior methods. 
Ren et al.~\cite{kgamd} built a knowledge graph from API documentation to model usage constraints like call order and condition checks, for detecting API misuses.
Li et al. \cite{Zhang-et-al} presented an improved API misuse detection method that integrates usage constraints from client code, documentation, and libraries to generate comprehensive AUGs.
Ma et al.~\cite{ma2024api} propose GraphiMuse, which encodes API usages as AUGs and learns probabilistic usage patterns by aggregating rules from code and documentation, representing each rule's trustworthiness as a probability.
Wang et al.~\cite{wang2023apicad} proposed APICAD,  which enhances API misuse detection by  inferring specifications from both code and documentation using static analysis and symbolic execution, targeting C++ programs at the LLVM IR level. 

Unlike these methods, \mugnn{} is an unsupervised GNN model that learns API usage patterns directly from real examples.
Our evaluation on the MUBench~\cite{amann2016mubench} dataset shows \mugnn{}’s effectiveness over prior methods~\cite{MuDetect, kgamd, Zhang-et-al, ma2024api} in terms of precision, recall, and F-score.
Further, unlike AUGs, which capture control and data flow, AFGs additionally model API call sequences with nodes as statements (not variables), enabling more precise representation of API usage behaviour.
Futhermore, as APICAD~\cite{wang2023apicad} does not support Java, a direct comparison with it is infeasible.

Kang and Lo~\cite{kang2021active} proposed Actively Learned Patterns (ALP), which frames API misuse detection as a classification task using active learning and human supervision to mine subgraphs that effectively distinguish correct from incorrect usages.
 In contrast, \mugnn{} employs an unsupervised GNN model to learn API usage patterns and detect misuses without manual intervention. (We could not directly compare with this work due to errors encountered while using the artifacts.)

\noindent{\it Domain-specific API misuse detection.} 
Recent work has explored domain-specific API misuse detection.
For example, LLMAPIDet~\cite{wei2024demystifying} 
detects misuses of deep learning (TensorFlow and PyTorch) APIs by applying few-shot prompting with a heavyweight LLM (ChatGPT~\cite{openai2023chatgpt}). 
Even on domain-specific test set, it detected only $48$ out of $291$ API misuses ($16.49$\%).
In contrast, AFGNN is a lightweight, semantics-aware, and language-agnostic graph-based model that learns API usage patterns directly from the AFGs, making them not directly comparable.
Further, Cryptolation~\cite{frantz2024methods} and CryptoGo~\cite{li2022cryptogo} detect  cryptographic API misuses in Python and Go, respectively, using static analysis and manually crafted rules, highlighting the prevalence of misuse patterns in security-critical code.
In contrast, AFGNN is a general, graph-aware, unsupervised approach that learns API usage patterns directly from real code, without relying on domain assumptions or handcrafted rules.

\section{\afgnn{} Framework}

We present \mugnn{}, a GNN-based framework that detects potential API misuse and recommends frequent ways of using an API based on open-source examples by using a new API Flow Graph (AFG) representation.
Figure~\ref{fig:mugnn-pipeline} shows the end-to-end workflow of our framework. 
As shown, our framework begins by generating the AFGs using an AFG generator for the method-level API usage examples (Section~\ref{s:api-flow-graph}).
Subsequently, our GNN model generates the graph embeddings from these AFGs (Section~\ref{s:mugnn-methodology}). 
These graph embeddings are then clustered (Section~\ref{s:api-usage-clustering}). 
The hypothesis is that the large clusters indicate frequent/common API usage patterns, while smaller clusters indicate infrequent usage or potential misuse~\cite{monperrus2013detecting, code-kernel, lindig2015mining, li2005pr, kang2021active}. 
These clusters can be used for different API-related tasks, such as recommending frequent usage patterns or detecting potential API misuse.


\subsection{API Flow Graph (AFG) Representation}
\label{s:api-flow-graph}

\begin{figure*}[t]
\centering
\includegraphics[width=0.8\textwidth]{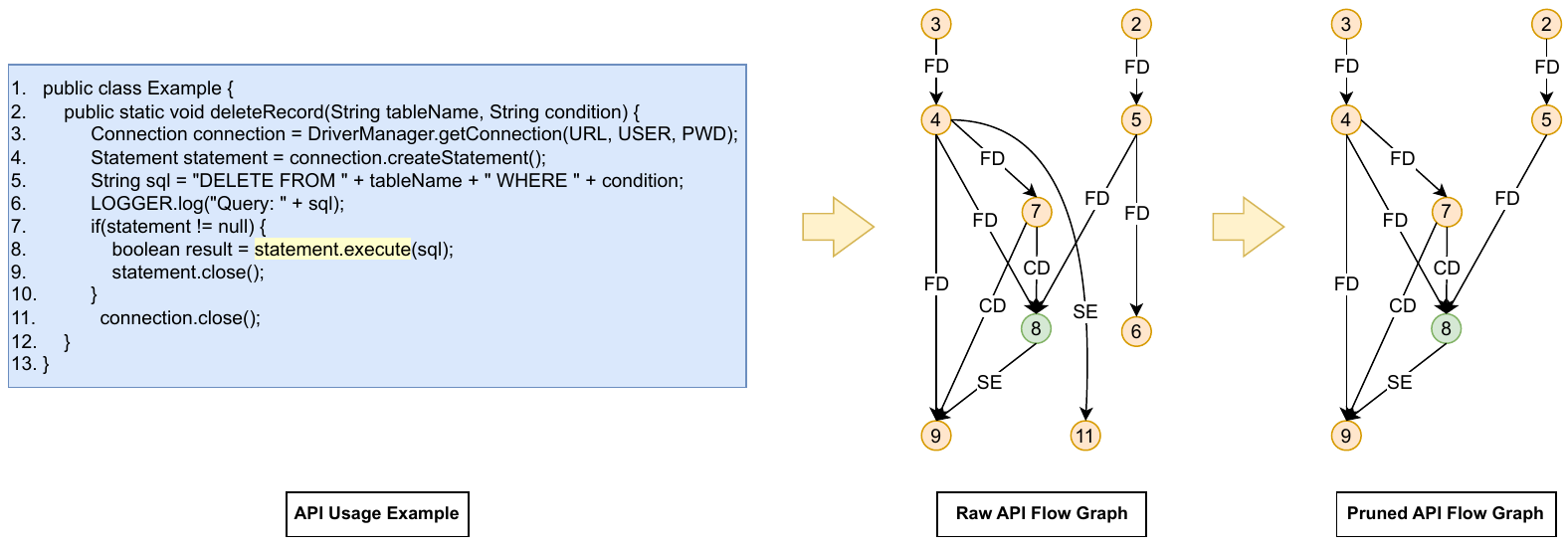}
\caption{Example of an API Flow Graph (AFG) for a {\tt Statement.execute()} usage. 
The graph captures data flow (FD), control flow (CD), and sequence (SE) edges among program lines. 
}
\label{fig:api-flow-graph-example-and-before-and-after-pruning}

\label{fig:api-flow-graph-example}
\Description{}
\end{figure*}

\subsubsection{AFG Definition}
An API Flow Graph (AFG) is a directed graph representation of a given API usage code snippet, defined as a three-tuple \( G = \langle V, E, X \rangle \). 
Here, $V$ is the set of vertices, with each vertex representing an embedding of a program line in the code,
$E$ is the set of edges denoting the dependencies between program lines, and
$X$ is the set of edge labels $\{FD, CD, SE\}$, representing data flow edges, control flow edges, and sequence edges. 
Figure~\ref{fig:api-flow-graph-example} shows an API usage example and its corresponding AFG with nine vertices (line numbers are shown instead of embeddings for simplicity) and $11$ edges. 
We describe the three types of edges below.

\noindent{\it Data flow edges (FD).} A data flow edge in AFG represents a data dependency between API-related operations in the code, indicating the flow of data from a $src$ node to a $target$ node.
For example, in Figure~\ref{fig:api-flow-graph-example}, the variable {\tt statement} at Line 4 stores the result of an API call {\tt createStatement()}, and  {\tt statement} is later used at Line 8 to invoke another API call {\tt execute()}. 
Hence, an FD edge connects the node representing Line 4 to the node representing Line 8.

\noindent{\it Control flow edges (CD).} A control flow edge represents a control dependency between API-related operations, indicating the flow of control (execution order) from a $src$ node to a $target$ node. 
The  $src$ node can be the starting line of a function, a branch condition of an \texttt{if} statement, a loop condition, a \texttt{switch} condition, or the beginning of a \texttt{try-catch} block.
The  $target$ node is a statement located within the bodies of these structural constructs.
These edges model instances where an API usage is preceded by verification checks (such as null pointer or array bounds checks) or protected by exception-handling constructs (like try-catch).
For example, in Figure~\ref{fig:api-flow-graph-example}, the API call {\tt execute()} at Line 8 is executed only if {\tt statement != null} at Line 7.
Hence, a CD edge connects the node representing Line 7 to the node representing Line 8.

\noindent{\it Sequence edges (SE).}
A sequence edge represents the sequential order in which API calls are made on a particular receiver object, capturing the correct sequence of operations to ensure proper API usage within a code snippet. 
SE edges are created automatically by linking each API call to its immediately following API call on the same object.
For example, in Figure~\ref{fig:api-flow-graph-example}, {\tt connection.createStatement()} at Line~4 is called before {\tt connection.close()} at Line~11, and both calls operate on the same {\tt connection} object.
Hence, the SE edge connects the node with Line~4 to the node with Line~11, but not Line 3  as the latter invokes an API on a different object ({\tt DriverManager}).
Similarly, in Figure~\ref{fig:br-readLine-usage}, 
a sequence edge connects the {\tt br.readLine()} and {\tt br.close()} calls to indicate the necessary order of execution, where the closing of the {\tt BufferedReader} ({\tt br.close()}) must follow the reading of a line ({\tt br.readLine()}), reflecting common practice, where resources are closed after use.




\subsubsection{AFG Generator}
\label{ss:afg-generator}
We implement an AST-based AFG generator for Java that constructs AFGs by capturing data flow, control flow, and API call sequences at the source-code level.
Our AST-based approach offers the advantage that it does not require compilable Java code, but it can generate AFG from any syntactically correct (partial) Java code, which allows us to consider any code snippet that contains API usage. 
AFG generator parses each Java code snippet into an AST and extracts statement-level nodes (e.g., variable declarations, assignments, API calls, conditionals).
Using the generated AST, FD edges are added by linking source nodes representing variable definitions to target nodes representing their corresponding uses.
CD edges are added from source nodes corresponding to  branch conditions in  structural constructs (e.g., conditionals, loops, and \texttt{try–catch} blocks) to target nodes representing statements within their bodies.
SE edges are created by linking each API call to its immediate successor API call on the same receiver object. 
To construct these edges, we identify API callsites (location where an API is invoked, e.g., {\tt statement.execute(sql)} is a callsite for {\tt execute()} API) and then analyze the surrounding statements to derive object-specific call sequences.


Figure~\ref{fig:api-flow-graph-example} shows the generated AFG for a sample API usage example.
In reality, every node in an AFG is an embedding of a program line in the code snippet.
We generate these embeddings by using CodeT5+~\cite{CodeT5+} in our approach (see Section~\ref{s:discuss}).

\subsubsection{AFG Pruning}
\label{ss:afg-pruning}

At inference time, given a targeted API of interest, a complete AFG is unnecessary as only the portion relevant to the API's usage is required. 
While the raw AFG (discussed above) captures the data flow, control flow, and API call sequence across code lines, it often includes extraneous nodes and edges that are unrelated to the specific usage of the targeted API, particularly in snippets containing multiple APIs. 
To address this, we apply a pruning algorithm on the raw AFG that retains only the subgraph relevant to the target API call. 
Specifically, it preserves nodes that are predecessors or successors of the API node, defined as those reachable from or to the API call through control, data, or sequence edges. 
The pruning algorithm further eliminates non-informative structures, such as class-level links, CD edges originating from method signatures, and self-loops, while merging nodes mapped to the same code line for compactness. 
The resulting subgraph compactly captures the target API's contextual usage. 
We further tested reversing or duplicating edges, but they showed no improvement and are disabled by default.

In Figure~\ref{fig:api-flow-graph-example}, if the target API method is \texttt{Statement.execute()}, then Line~6 and Line~11 are not relevant to API usage because Line~6 is only used for logging, and Line~11 closes the connection object, and they are not connected to the {\tt Statement.execute()} call (node 8). 
While \texttt{createStatement()} at Line 4 and \texttt{close()} at Line 11 are relevant for analyzing the usage of the \texttt{Connection.close()} API method, it is outside the scope when \texttt{Statement.execute()} is the targeted API under investigation. 
Since our focus is on detecting misuse patterns for a specific API, we consider one API misuse at a time during inference, even in real-world scenarios where multiple APIs appear within a snippet.
Therefore, as the nodes numbered 6 and 11 are not reachable to/from the API call node (node 8), they 
are removed from the final AFG as part of the pruning process.

We evaluated AFG pruning on a dataset containing multiple APIs alongside the targeted API, and found it effective in multi-API settings. 
This is because it removes nodes and edges related to non-target APIs while preserving only statements that influence the targeted API call, thereby reducing noise from unrelated API usage.
This design naturally scales to complex code contexts common in real-world scenarios, where its benefits in focusing on the targeted API are expected to be even greater. 
We apply the AFG pruning process only during evaluation/inference, but not during the pre-training of AFGNN, as it requires knowing the target API in advance and is thus impractical for the pre-training stage.
During pretraining, AFGNN instead uses a generic context prediction objective on the raw AFG.

\begin{figure*}[htp]
\centering
\begin{minipage}{0.48\linewidth}
    \centering
    \includegraphics[trim={0cm 0cm 0cm 0cm}, clip, width=\linewidth]{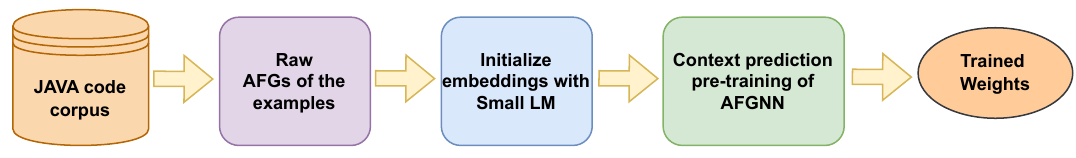}
    \caption{\mugnn{} pre-training pipeline}
    \label{fig:mugnn-pretraining-pipeline}
\end{minipage}
\hfill
\begin{minipage}{0.48\linewidth}
    \centering
    \includegraphics[width=\linewidth]{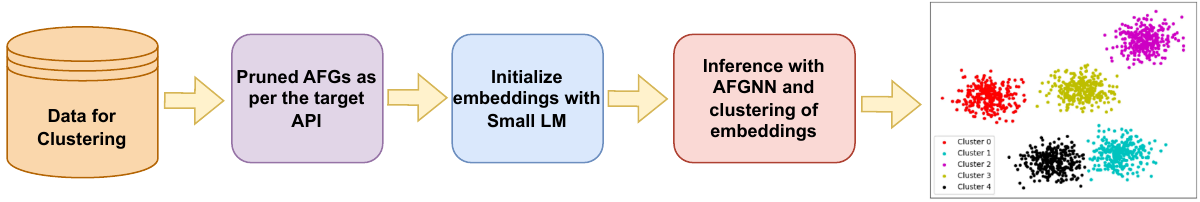}
    \caption{\mugnn{} clustering pipeline}
    \label{fig:mugnn-clustering-pipeline}
\end{minipage}
\Description{}
\end{figure*}



\subsection{\mugnn{} Model}
\label{s:mugnn-methodology}

Graph Neural Networks (GNN) are widely used to obtain meaningful representations of nodes or entire graphs. 
We have used GNN to get the embedding from the API Flow Graph (AFG) representation. 
This section explains how we pre-train our \mugnn{} model.

\subsubsection{\mugnn{} Architecture}
\label{s:mugnn-model-architecture}

\mugnn{} can support standard Graph Neural Network (GNN) architectures, and we empirically identified the most effective GNN architecture for API misuse detection. 
\mugnn{} uses the AFG representation of code and learns the node representations through message-passing, where nodes iteratively exchange and aggregate information from their neighbours.
A $k$-layer GNN can aggregate the node embeddings up to $k$-hops, with the message-passing function defined by its architecture.
For \mugnn{}, we have experimented with four types of GNN architectures: GCN~\cite{kipf2016semi}, GAT~\cite{velivckovic2017graph},  RGCN~\cite{thanapalasingam2022relational}, and RGAT~\cite{busbridge2019relational} (explained in Section~\ref{s:back}).
We present in this paper the results using GCN and RGCN architectures in the \mugnn{} model due to their superior performance over GAT and RGAT.
A five-layer GNN architecture is sufficient for capturing the complexity of our AFGs because the pruned graphs in our dataset are small, with edges ranging from 1 to 61, allowing information to propagate across all nodes reachable from the API callsite.


After the message passing step, we obtain the final set of node embeddings for the input AFG. 
Standard pooling functions like mean, max, and min, though common, are not optimal for capturing API flow information, as they lose API-specific details (based on our experiments).
Using the embedding of the API callsite (API node) as the graph representation yielded better results because message passing centering the API call captures its neighbourhood with API-related operations.
Hence, we present the results using the \mugnn{} embedding of the API node.
For code examples with multiple callsites  having the same API call, we apply mean pooling to the API nodes' embeddings to obtain the final API flow embedding.


\subsubsection{\mugnn{} Pre-training Dataset} 
\label{ss:pre-training-data}

Pre-training trains a model on a large dataset in an unsupervised or self-supervised way to learn general code patterns,
and we use the pre-trained \mugnn{} model later for inference and generating graph embeddings, without fine-tuning. 
Since our downstream task is unsupervised clustering, effective API-usage clustering through transfer learning requires pre-training on a large and relevant dataset.

For \mugnn{}, we pre-train on the java-large~\cite{java-large-dataset} dataset from Code2Seq~\cite{alon2018codeseq}, sourced from $9,500$ top-starred GitHub Java projects 
created since January 2007 that have active software development histories. 
We extracted method-level code examples from the raw project-level data, focusing only on those that use popular Java library APIs for our downstream API-usage clustering tasks.
We focus on popular Java APIs because they are widely used in real-world software projects, enabling the model to learn representative and generalizable API usage patterns, consistent with prior empirical studies~\cite{MuDetect,Zhang-et-al,ma2024api, kgamd} on the MuBench benchmark.
In total, we collected approx. 1.5 million unique Java API usage examples that are split into training, testing, and validation data sets in an 8:1:1 ratio. 
For pretraining, we considered AFGs with at least three edges to ensure informative graph structures and support effective and unbiased model learning.

Our pre-training dataset includes diverse code samples from various domains (identified through manual inspection) such as Networking, Concurrency and Multithreading, Database Operations, File and I/O Operations, Date and Time Management, Resource Management and Class Loading, String and URI Handling, Graphics, System and Memory Management, and GUI Operations.
This broad exposure helps \mugnn{} capture diverse API usage patterns, enhancing its ability to generalize,
and accurately evaluate API usage.

\subsubsection{\mugnn{} Pre-training}
\label{s:mugnn-pre-training}
For pre-training, we first generate AFGs from the Java method-level examples. 
Since GNNs require dense node representations as input, we can initialize the nodes with embeddings from small LMs such as CodeT5+~\cite{CodeT5+}, CodeBERT~\cite{feng-etal-2020-codebert}, and UniXcoder~\cite{Unixcoder}.
We use CodeT5+ for initialization, as it provides the best overall performance. 
Our pre-training experiment on the context-prediction task show that CodeT5+ achieves an accuracy of $0.902$, outperforming UniXcoder ($0.884$) and CodeBERT ($0.786$), which is why we select CodeT5+ to initialize node representations.


Figure~\ref{fig:mugnn-pretraining-pipeline} illustrates the pre-training pipeline of \mugnn{}.
 We pre-train \mugnn{} using the context prediction objective~\cite{Hu2020Strategies}, where subgraphs are used to predict surrounding graph structures, aiming to map nodes in similar structural contexts to nearby embeddings.
 This objective naturally aligns with our goal to get meaningful embeddings for API nodes, ensuring that API nodes of similar API usage examples have similar embeddings as they would appear in a similar context.
We have not pruned the AFGs during pre-training as context-prediction pre-training is unsupervised and not specific to API usage.
Thus, \mugnn{} leverages unsupervised context-prediction pretraining to generate flow-sensitive code embeddings, which are then clustered (see Figure~\ref{fig:mugnn-clustering-pipeline}) to identify API usage patterns and detect potential misuse, enabling independence from labeled data and offering flexibility in analyzing unfamiliar API usages.

\subsection{API Usage Clustering}
\label{s:api-usage-clustering}

\begin{algorithm}[t]
\SetKwFunction{clustering}{clusterTheEmbeddings}
\SetKwProg{myproc}{Procedure}{}{}
\SetKwInput{KwInput}{Input}                
\SetKwInput{KwOutput}{Output}              
\KwInput{E: Embeddings as $[e_1, e_2, \ldots, e_n]$, where $e_i$ is the \mugnn{} embedding for $i$th example; M: Birch clustering algorithm}
\KwOutput{C: Cluster labels as $[c_1, c_2, \ldots, c_n]$, where $c_j$ is the predicted cluster label for jth example}
\myproc{\clustering{$E, M$}  }{
$min\_db\_score \gets \infty$\;
$C \gets [0, 0, \ldots, 0]$ \tcp*{Initially, all examples are in the 0th cluster}
\For{$cluster\_cnt \gets 2 \;\KwTo \;n$}{
    $pred\_labels \gets M(E, \;cluster\_cnt)$\; \label{algo:query-M}
    $db\_score \gets calcDBscore(E, \;pred\_labels)$\;
    \If{$db\_score < min\_db\_score$}
    {
        $min\_db\_score \gets db\_score$\;
        $C \gets pred\_labels$ \tcp*{Clustering improved}
    }
}
}
\caption{Algorithm to find the best clustering}
\label{algo:clustering-algo}
\Description{}
\end{algorithm}

We use the pre-trained \mugnn{} to generate embeddings for API usage examples and then cluster them using the popular BIRCH algorithm~\cite{zhang1997birch} to identify API misuses (as shown in Figure~\ref{fig:mugnn-clustering-pipeline}).
We chose BIRCH due to its hierarchical and memory-efficient design, incremental clustering capability, and practical speed advantages, as well as its flexibility in fine-tuning via the threshold parameter. 
Although our evaluation dataset is relatively small, BIRCH demonstrated competitive performance and interpretability compared to alternative clustering algorithms reported in the literature. 
We report results using the optimal thresholds for \mugnn{} and baseline models for the API usage clustering.

The optimal number of clusters depends on factors such as the specific API and example count.
To determine the best clustering, we use the Davies-Bouldin (DB)~\cite{davies-bouldin} metric, selecting the number of clusters that yield the best DB score, with lower scores indicating better clustering.
\begin{equation}
DB = \frac{1}{K} \sum_{i=1}^{K} \max_{j \neq i} \left( \frac{S_i + S_j}{M_{ij}} \right)
\end{equation}
where $K$ is the number of clusters, $S_i$, and $S_j$ denote the average distances of points in clusters $i$, and $j$ to their respective centroids, $M_{ij}$ is the distance between cluster centroids of clusters $i$ and $j$.
The algorithm in Figure~\ref{algo:clustering-algo} is used to determine the best clustering for a particular API.
Initially, all the examples are considered to be in
a single cluster, and iteratively clustering the examples for all possible cluster counts (from \texttt{2} to the total number of examples, \texttt{n}), calculating the DB score in each iteration and
selecting the 
final clustering with the minimum DB score.

\section{Experimental Methodology}

We evaluate \mugnn{} on the following research questions.

\noindent{\textbf{RQ1.}} How effective are the \mugnn{} model embeddings at clustering different API usage patterns?\\
\noindent{\textbf{RQ2.}} How effectively does \mugnn{} detect API misuse in real-world examples (written by developers) from MUBench~\cite{amann2016mubench} dataset?\\
\noindent{\textbf{RQ3.}} How do addition of sequence edges help \mugnn{} in identifying API usage patterns?\\
\noindent{\textbf{RQ4.}} How does AFG pruning (removing nodes and edges not related to the target API usage) affect \mugnn{} performance?\\

\noindent{{\bf Experimental setup.}
We conducted all our experiments on a machine with an Intel Xeon Gold 5326 CPU (32 cores, 2.90 GHz) and a 1.41 GHz NVIDIA Ampere A100 GPU with 80 GB global memory, which was used for training and inference of \mugnn{}.
We trained the \mugnn{} model with the context-prediction pre-training objective, using a learning rate of $5e^{-5}$, a batch size of 256, and the Adam optimizer ($\epsilon = 1e^{-8}$) until validation accuracy plateaued for five consecutive epochs.
We implemented the \mugnn{} pipeline in Python (v3.12.2) with PyTorch (v2.2.1). 
We present the performance results of two variants of \mugnn{}: 5-layer GCN and RGCN (denoted as \mugcn{} and \murgcn{}), both pre-trained with the same setup. 
As baselines, we consider UnixCoder and GraphCodeBERT, two state-of-the-art small LMs (criteria explained in Section~\ref{s:discuss}), and have utilized the same system for inference. 
For all the baseline models, we used pre-trained weights from HuggingFace~\cite{huggingface-link}. 
Since very large LLMs (e.g., GPT-4, DeepSeek) are computationally expensive and hard to fine-tune, we restrict our baselines to lightweight small LMs and graph-based approaches that can be trained and evaluated efficiently within our setup.
Further, a comparison with CodeKernel~\cite{code-kernel} is not feasible as it is not open-source, and we contacted the authors, who confirmed that the code is currently unavailable.
All our datasets and source code are publicly available~\cite{suppl}.

\noindent{\bf {Metrics Used.}} 
To determine the optimal number of
clusters for a particular API  and the clustering quality, we use an intrinsic evaluation measure Davies-Bouldin~\cite{davies-bouldin} metric (see Section~\ref{s:api-usage-clustering}).
To compare the clusters produced by \mugnn{} with those from the baselines, we used two popular external clustering index metrics: Rand Index (RI)~\cite{rand-index} and Mutual Information (MI)~\cite{mutual-info-score}, which measure the similarity between the predicted clusters and ground truth. 
We use the ``Adjusted” versions of these metrics from scikit-learn library~\cite{sklearn-website}, as they penalize random labeling and large numbers of clusters. 
All clustering metrics are sourced from the scikit-learn library~\cite{sklearn-website}.
The RI (eq.~\ref{eq:RI}) measures the proportion of sample pairs that are either assigned to the same cluster in both the predicted and ground truth labels, or assigned to different clusters in both. 
\begin{equation}
RI = \frac{\text{Number of agreements}}{\text{Total number of pairs}}
\label{eq:RI}
\end{equation}
The Adjusted Rand Index (ARI) measures clustering similarity by adjusting the observed agreement between two clusterings to account for agreement expected by chance.
ARI ranges from -1 (poor agreement) to 1 (perfect match), with 0 indicating random labeling.

The MI measures the amount of shared information between two clusterings. It is given by:
\begin{equation}
MI(U, V) = \sum_{u \in U} \sum_{v \in V} P(u, v) \log \left( \frac{P(u, v)}{P(u)P(v)} \right)
\label{eq:MI}
\end{equation}
where \( P(u) \) and \( P(v) \) are the probabilities of clusters $u$  and $v$ in $U$ (the true labels) and $V$ (the predicted labels), respectively, and \( P(u, v) \) is their joint probability.


Adjusted Mutual Information (AMI) normalizes mutual information by correcting for the agreement expected by chance, using the entropies of the two clusterings, and ranges from 0 (no agreement beyond chance) to 1 (perfect agreement).

\section{Dataset}
\label{s:dataset}
This section 
describes the two datasets used in our evaluation: 
(1)~a new labeled dataset based on  CodeKernel~\cite{code-kernel} data for API usage clustering evaluation (RQ1) and ablation studies (RQ3 and RQ4), and 
(2) the API misuse dataset from MUBench for comparison with baselines (RQ2), respectively.




\subsection{API Usage Clustering}
\label{ss:api-usage-clustering-data}

Due to the absence of real-world expert-annotated API usage clusters, we curate a new dataset based on CodeKernel ~\cite{code-kernel},
which provides usage examples for a set of $25$ popular Java APIs via its GitHub page~\cite{code-kernel-dataset}, with the number of examples per API ranging from $6$ to $1,368$. 
Among them, we consider a total of $21$ APIs (listed in Table~\ref{table:external-clustering-evaluation}) that had at least $30$ examples.
For each selected API, we randomly sampled and manually labeled around $30$ examples (excluding very small snippets to ensure meaningful clustering) to keep human annotation feasible, maintain consistency across APIs and avoid possible dataset imbalance.
\begin{table}[t]
\caption{Sample rules for manual API usage clustering}

\label{tab:manual-clustering-rules}
\centering
\footnotesize
\begin{tabular}
{p{0.03\linewidth}p{0.87\linewidth}}
\toprule
\textbf{Rule} & \textbf{Description} \\
\midrule
R1 & API calls with different signatures (distinct argument types due to inheritance/polymorphism) result in separate clusters. \\
R4 & API usage enclosed within \texttt{if-else} or \texttt{try-catch} blocks leads to different clusters. \\
R5 & If the result of an API call is assigned to, or appended to a variable (e.g., via append() or add()), it belongs to a separate cluster. \\
R6 & Multiple calls to the API within loops or conditional statements should be clustered differently. \\
R10 & API calls within nested conditionals or nested loops should form separate clusters. \\
\bottomrule
\end{tabular}
\end{table}

\begin{figure}[t]
\centering
\begin{subfigure}[b]{0.45\textwidth}
\centering
\begin{lstlisting}[language=Java, numbers=left, numberstyle=\tiny, stepnumber=1, numbersep=3pt, xleftmargin=2em, breaklines=true, breakatwhitespace=true, basicstyle=\ttfamily\footnotesize]
getResourceStreamWithClassLoader(ClassLoader classLoader,String path){
    if (classLoader != null) {
      URL url = classLoader.getResource(path);
      if (url != null) {
        return new UrlResourceStream(url);
      }
    }}
\end{lstlisting}
\caption{API result is assigned to a variable}
\label{fig:rule1-callable}
\end{subfigure}
\hfill
\begin{subfigure}[b]{0.45\textwidth}
\centering
\begin{lstlisting}[language=Java, numbers=left, numberstyle=\tiny, stepnumber=1, numbersep=3pt, xleftmargin=2em, breaklines=true, breakatwhitespace=true, basicstyle=\ttfamily\footnotesize]
findCurrentResourceVersion(String resourceUrl){
        ClassLoader cl = getClass().getClassLoader();
        return cl.getResource(resourceUrl);
}
\end{lstlisting}
\caption{API result is directly returned without assignment}
\label{fig:rule1-runnable}
\end{subfigure}
\caption{Examples for Rule R5. Although both examples call \texttt{getResource()}, they differ in how the result is used.
}
\label{fig:rule1}
\Description{}
\end{figure}

\begin{figure*}[htp]
\centering
\begin{subfigure}[t]{0.5\textwidth}
\centering
\begin{lstlisting}[style=java-code-style, numbers=left, numberstyle=\tiny, stepnumber=1, 
xleftmargin=1em, numbersep=3pt, breaklines=true, breakatwhitespace=true, basicstyle=\ttfamily\scriptsize]
void pattern(Map foregroundDomainMarkers, ..., Marker marker) {
  ArrayList markers = (ArrayList) foregroundDomainMarkers.get(...);
  if (markers != null) {
    markers.remove(marker);
  }
}
\end{lstlisting}
\begin{lstlisting}[basicstyle=\ttfamily\tiny, breaklines=true]
Line_1 $$ void pattern(...) { --> Line_2 $$ ArrayList markers = ... [FD]
Line_2 $$ ArrayList markers = ... --> Line_3 $$ if (markers != null) { [FD]
Line_3 $$ if (markers != null) { --> Line_4 $$ markers.remove(marker); [CD]
Line_2 $$ ArrayList markers = ... --> Line_4 $$ markers.remove(marker); [FD]
Line_1 $$ void pattern(...) { --> Line_4 $$ markers.remove(marker); [FD]
\end{lstlisting}
\caption{Correct use example from MUBench with its AFG.}
\label{fig:mubench-correct}
\end{subfigure}
\hfill
\begin{subfigure}[t]{0.48\textwidth}
\centering
\begin{lstlisting}[style=java-code-style, numbers=left, numberstyle=\tiny, stepnumber=1, 
xleftmargin=2em, numbersep=3pt, breaklines=true, breakatwhitespace=true, basicstyle=\ttfamily\scriptsize]
void pattern(Map foregroundDomainMarkers, ..., Marker marker) {
  ArrayList markers = (ArrayList) foregroundDomainMarkers.get(...);
  markers.remove(marker);
}
\end{lstlisting}
\begin{lstlisting}[basicstyle=\ttfamily\tiny, breaklines=true]
Line_1 $$ void pattern(...) { --> Line_2 $$ ArrayList markers = ... [FD]
Line_1 $$ void pattern(...) { --> Line_3 $$ markers.remove(marker); [CD]
Line_1 $$ void pattern(...) { --> Line_3 $$ markers.remove(marker); [FD]
Line_2 $$ ArrayList markers = ... --> Line_3 $$ markers.remove(marker); [FD]
\end{lstlisting}
\caption{Misuse example from MUBench with its AFG.}
\label{fig:mubench-misuse}
\end{subfigure}
\caption{Examples of correct use and misuse from MUBench, shown with their corresponding AFGs.}
\label{fig:mubench-examples}
\Description{}
\end{figure*}

To generate high-quality labels, we enlisted two Computer Science graduate students, each with over two years of Java development experience. 
To minimize subjective bias, they first independently reviewed a subset of examples and collaborated to derive a unified set of 14 rules for labeling the dataset.
They first annotated a subset of examples separately, compared their labels, and refined the rules iteratively to resolve disagreements.
After reaching consensus, the annotators re-labeled the dataset using the finalized rules, resulting in a consistent and reliable ground-truth dataset.

We show $5$ representative rules in Table~\ref{tab:manual-clustering-rules} (among $14$, due to space limitations). 
The rules cover factors influencing clustering decisions, such as variations in API signatures, parameter scopes, control-flow contexts, and other usage patterns. 
As illustrated in Figure~\ref{fig:rule1}, although both examples invoke the same API \texttt{getResource()}, they differ in how the result of the API call is handled. 
In the first example, the result is assigned to a variable for further checking, while in the second example, the API result is returned directly without any intermediate handling.
As per Rule R5, such semantic differences lead to separate clusters, as they reflect distinct API usage patterns.
The full rule set is in the supplementary material (Appendix A and B)~\cite{suppl}.
Using these rules, annotators independently clustered the examples. 
Further, they cross-verified each other's annotations and resolved disagreements using mutual discussions. 
This led to a consistent set of labelling rules, which can be extended to other APIs. 
The curated dataset was used solely for testing, not training, in our RQ1 evaluation.
\subsection{MUBench dataset}
\label{ss:mubench-data}
MUBench~\cite{amann2016mubench} is a popular dataset for API misuse detection, comprising 162 correct usage examples and corresponding misuse descriptions from 67 open-source Java projects. 
Since it is built from real-world open-source Java projects, MUBench offers a reliable and practical dataset for evaluating API misuse detection.
The dataset includes correct usage examples, supplemented by YML files detailing misuse cases, file URLs, descriptions, and proposed fixes.
Due to the dataset’s age, some repositories were inaccessible, so we collected all available examples and YML files and reconstructed the missing misuse cases based on the YML descriptions.
Further, since \mugnn{} operates at the method level, we focused on extracting only the methods associated with API misuse. 
As a whole, we compiled a total of $324$ examples: $162$ correct usage examples downloaded directly from the MUBench repository and $162$ corresponding misuse examples.
Figure \ref{fig:mubench-examples} shows examples of correct use and misuse from MUBench alongside their corresponding AFGs. 
In the correct case, the conditional check {\tt(if (markers != null))} is preserved in the AFG, ensuring safe API usage. 
In contrast, the misuse omits this check, and the resulting AFG lacks the control-dependency edge, highlighting how AFGs capture semantic differences between correct and incorrect API usages.

\section{Results}
\label{s:eval}

\subsection{RQ1: Effectiveness of \mugnn{}} 
\label{s:mugnn-evaluation}

\mugnn{} embeddings are evaluated for their effectiveness in clustering similar API usages, using labeled examples from 21 popular Java APIs (see Section~\ref{ss:api-usage-clustering-data}).
Table~\ref{table:external-clustering-evaluation} presents the clustering results for four different models (\mugcn{}, \murgcn{}, and the two baseline models), showing the Rand Index (RI) and Mutual Information (MI) scores based on predicted clusters and ground-truth.
In each row, the highest RI and MI scores are indicated in bold. 
The suffix ``Birch~(1.5)” in the first row indicates the threshold of 1.5 used in the BIRCH clustering algorithm to control the maximum diameter of subclusters (similarly, 2.5).

\begin{table*}[t]
\caption{External index results for API usage clustering. The upward arrow indicates that higher RI and MI scores are better. The highest MI scores and RI scores are highlighted in each row}
\label{table:external-clustering-evaluation}
\centering
\small
\resizebox{0.85\textwidth}{!}{\scriptsize
\begin{tabular}{lrrrrrrrr}
\toprule
& \multicolumn{2}{c}{\textbf{UnixCoder-Birch (1.5)}} & \multicolumn{2}{c}{\textbf{GraphCodeBERT-Birch (1.5)}} & \multicolumn{2}{c}{\textbf{\mugcn{}-Birch (1.5)}} & \multicolumn{2}{c}{\textbf{\murgcn{}-Birch (2.1)}}\\
 \cmidrule(lr){2-9} 
\textbf{API methods} & \multicolumn{2}{c}{\textbf{126M}} & \multicolumn{2}{c}{\textbf{124M}} & \multicolumn{2}{c}{\textbf{331K}} & \multicolumn{2}{c}{\textbf{1.3M}}\\
 \cmidrule(lr){2-9} 
& \textbf{RI score $\uparrow$} & \textbf{MI score} $\uparrow$ & \textbf{RI score} $\uparrow$ & \textbf{MI score} $\uparrow$ & \textbf{RI score} $\uparrow$ & \textbf{MI score} $\uparrow$ & \textbf{RI score} $\uparrow$ & \textbf{MI score} $\uparrow$\\
\midrule
Classloader.getResource() & -0.004 & -0.005&\textbf{0.401}&\textbf{0.454}&0.254&0.23&0.322&0.396\\
\midrule
Thread.start()&-0.003&-0.003&0.158&0.176&0.114&0.231&\textbf{0.439}&\textbf{0.472}\\
\midrule
Statement.execute()&0.219&0.235&0.237&0.33&\textbf{0.34}&\textbf{0.422}&0.174&0.184\\
\midrule
BufferedReader.read()&0.027&0.055&0.299&0.366&0.144&0.238&\textbf{0.412}&\textbf{0.492}\\
\midrule
Timestamp.compareTo()&\textbf{0.497}&\textbf{0.497}&0.239&0.239&0.39&0.489&0.325&0.325\\
\midrule
DataInputStream.readLine()&0.163&0.182&0.4&0.416&0.383&0.423&\textbf{0.41}&\textbf{0.465}\\
\midrule
ServerSocket.bind()&0.114&0.141&\textbf{0.288}&\textbf{0.34}&0.098&0.198&0.231&0.298\\
\midrule
ExecutorService.submit()&\textbf{0.665}&\textbf{0.665}&0.192&0.205&0.021&0.056&0.278&0.306\\
\midrule
URI.getFragment() & 0.302 & 0.302 & 0.449 & 0.538 & 0.433 & 0.449 & \textbf{0.683} & \textbf{0.719} \\
\midrule
Calendar.getTime() & 0.219 & 0.235 & 0.304 & 0.423 & 0.064 & 0.207 & \textbf{0.313} & \textbf{0.425} \\
\midrule
Socket.connect() & 0.302 & 0.335 & \textbf{0.608} & \textbf{0.671} & 0.198 & 0.412 & 0.58 & 0.621 \\
\midrule
JPanel.add() & 0.114 & 0.157 & 0.273 & 0.352 & 0.055 & 0.122 & \textbf{0.384} & \textbf{0.49}\\
\midrule
FileChannel.read() & -0.004 & -0.004 & \textbf{0.171} & \textbf{0.182} & -0.007 & -0.007 & 0.142 & 0.157\\
\midrule
DateFormat.format() & -0.004 & -0.004 & 0.141 & 0.189 & 0.042 & 0.096 & \textbf{0.364} & \textbf{0.339}\\
\midrule
ClassLoader.loadClass() & -0.004 & -0.004 & 0.087 & 0.11 & -0.041 & -0.072 & \textbf{0.208} & \textbf{0.232}\\
\midrule
Runtime.freeMemory() & -0.004 & -0.008 & 0.143 & \textbf{0.246} & 0.007 & 0.059 & \textbf{0.151} & 0.175\\
\midrule
Graphics2D.fill() & \textbf{0.139} & \textbf{0.171} & 0.004 & 0.02 & 0.013 & 0.071 & 0.004 & 0.02\\
\midrule
DriverManager.getConnection() & 0.178 & 0.189 & 0.288 & \textbf{0.352} & 0.284 & 0.328 & \textbf{0.301} & 0.335\\
\midrule
URL.openConnection() & -0.004 & -0.004 & 0.408 & 0.476 & 0.357 & 0.379 & \textbf{0.506} & \textbf{0.578}\\
\midrule
File.toURI() & -0.004 & -0.004 & -0.01 & -0.011 & -0.02 & -0.038 & \textbf{0.171} & \textbf{0.171}\\
\midrule
BufferedReader.readLine() & 0.087 & 0.116 & 0.18 & 0.247 & \textbf{0.216} & 0.284 & 0.183 & \textbf{0.304}\\
\midrule
\textbf{Average} & 0.143 & 0.154 & 0.25 & 0.301 & 0.159 & 0.218 & \textbf{0.313} & \textbf{0.357}\\
 \bottomrule
\end{tabular}
}
\end{table*}

\begin{figure}[t]
\centering
\begin{lstlisting}[style=java-code-style, numbers=left, numberstyle=\tiny, stepnumber=1, xleftmargin=2em, numbersep=3pt, escapechar=|, breaklines=true, breakatwhitespace=true, basicstyle=\ttfamily\scriptsize]
public void paint(Graphics2D g2d){
    g2d.setColor(new Color(96, 96, 96));
    g2d.fill(area);
}
\end{lstlisting}
\caption{A short usage example of {\tt Graphics2D.fill()} API.}
\label{fig:graphics2d-fill-example}
\Description{}
\end{figure}


\begin{table}[t]
\caption{Comparison of UnixCoder, GraphCodeBERT, and \mugnn{} with statistical measures}
\label{table:comparison-evaluation}
\small
\centering
\begin{tabular}{lrrrr}
 \toprule
\textbf{Model Pairwise} & \multicolumn{2}{c}{\textbf{P-Value}} & \multicolumn{2}{c}{\textbf{P-Value (BH)}} \\
\midrule
\textbf{Comparisons} & \textbf{RI $\downarrow$} & \textbf{MI $\downarrow$} & \textbf{RI $\downarrow$} & \textbf{MI $\downarrow$} \\
\midrule
UnixCoder and \murgcn{} & 0.0022 & 0.0008 & 0.0067 & 0.0032 \\
\midrule
UnixCoder and \mugcn{} & 0.8938 & 0.1944 & 0.8938 & 0.2121  \\
\midrule
GraphCodeBERT and \murgcn{} & 0.0104 & 0.0227 & 0.0179 & 0.0302 \\
\midrule
GraphCodeBERT and \mugcn{} & 0.0053 & 0.0128 & 0.0129 & 0.0191 \\
\bottomrule
\end{tabular}
\end{table}


The last row of Table~\ref{table:external-clustering-evaluation} reports the average RI and MI scores, where \murgcn{} outperforms all models.
The closest is GraphCodeBERT, which is nearly $\text{95}$ times larger than \murgcn{}. 
Thus, \mugnn{} is not only more effective than state-of-the-art {small} LMs, but it is also more efficient and less resource-intensive. 
GraphCodeBERT benefits from data flow modeling, but it still lags behind \mugnn{}, which additionally captures API-centric control flow and call sequences.
Some APIs have very short examples, where token-based models such as UnixCoder and GraphCodeBERT perform well. Figure~\ref{fig:graphics2d-fill-example} illustrates such a case for the {\tt Graphics2D.fill()} API. 
While AFGNN relies on graph structure, the corresponding AFG in this setting contains only a few edges, thereby limiting contextual information.
Nevertheless, p-value analysis (discussed next) confirms that AFGNN’s performance gains are statistically significant, and its $100\times$ smaller size highlights its practical value.
UnixCoder has performed the best for three APIs {\tt Graphics2D.fill()}, {\tt Timestamp.compareTo()} and {\tt ExecutorService.submit()} where the code examples are typically small (2-3 lines in some cases) and textual patterns matter more than the data and control flow. 
\lstset{style=java-code-style}


\noindent \textbf{\textit{P-Value evaluation.}} 
\mugnn{} performs well in most API clusterings, as indicated by the RI and MI scores in Table~\ref{table:external-clustering-evaluation}. 
However, some scores for GraphCodeBERT and \mugnn{}, as well as UnixCoder and \mugnn{}, are quite close, indicating less difference between them. 
To understand the statistical significance of these results, we conducted a p-value evaluation, using pairwise comparisons and employing a significance threshold of $0.05$, and the results are summarized in Table~\ref{table:comparison-evaluation}.
Given the multiple pairwise statistical tests, we applied the Benjamini-Hochberg (BH) correction to control the Type-I error rate, with the adjusted p-values shown in the fourth and fifth columns of Table~\ref{table:comparison-evaluation}.
For our evaluation, the null hypothesis stated that there is no variation in the models' performances. 
Significant differences exist between UnixCoder and \murgcn{} (raw p-values of $0.0022$ for RI and $0.0008$ for MI, with BH-corrected p-values of $0.0067$ and $0.0032$, respectively) as well as between GraphCodeBERT and \murgcn{} (raw p-values of $0.0104$ for RI and $0.0227$ for MI, with BH-corrected p-values of $0.0179$ and $0.0302$). 
These differences are accompanied by large effect sizes, with Cohen’s $d$ values up to $-0.86$ and Cliff’s $\delta$ exceeding $-0.6$, indicating that the improvements of \murgcn{} are not only statistically significant but also practically meaningful. 
A p-value of less than $0.05$ allows us to reject the null hypothesis, highlighting the superior performance of \murgcn{}, particularly in comparison to UnixCoder and GraphCodeBERT.

On the other hand, there is no statistically significant or practically meaningful difference between UnixCoder and \mugcn{}, as reflected by high p-values (BH-corrected p-values of $0.8938$ for RI and $0.2121$ for MI) and negligible effect sizes (Cohen’s $d$ close to $0$ and small Cliff’s $\delta$ values).  
In addition, GraphCodeBERT performs better than \mugcn{} (with a large effect size). The reason is that \mugcn{} does not differentiate between FD, CD, and SE, which makes it less expressive, although \murgcn{} achieves the best overall performance.

\subsection{RQ2: API Misuse Detection on MUBench}
\label{ss:real-world-examples}

We evaluate \mugnn{} on the MUBench\cite{amann2016mubench} dataset to assess its effectiveness on real-world benchmarks.
We first compare \mugnn{} with GraphCodeBERT, which performs best among the small LM baselines, as shown in RQ1. 
Table~\ref{table:mubench-evaluation} summarizes the evaluation results on MUBench. 
We have used the RGCN variant of \mugnn{} (\murgcn{}) as it performed the best.
\begin{table}[t]
\caption{Misuse detection on MUBench}
\label{table:mubench-evaluation}
\centering
\small
\resizebox{\columnwidth}{!}{
\begin{tabular}{@{~}l@{~}c@{~}c@{~}c@{~}c@{\hskip 0.3cm}c@{~}c@{~}c@{~}c@{~}}
\toprule
\multirow{2}{*}{\textbf{Threshold}} & \multicolumn{4}{c}{\textbf{GraphCodeBERT-Birch (3.0)}} & \multicolumn{4}{c}{\textbf{\murgcn{}-Birch (3.0)}}\\
\cmidrule(lr){2-9} 
& \textbf{\footnotesize Accuracy} & \textbf{\footnotesize Precision} & \textbf{\footnotesize Recall} & \textbf{\footnotesize F1-score} & \textbf{\footnotesize Accuracy} & \textbf{\footnotesize Precision} & \textbf{\footnotesize Recall} & \textbf{\footnotesize F1-score}\\
\midrule
5\%	& \textbf{0.573} & 0.65 & 0.317 & 0.426 & 0.547 & 0.557 & 0.487 & 0.52 \\
\midrule
10\% & \textbf{0.573} & 0.636 & 0.341 & 0.444 & \textbf{0.584} & 0.572 & 0.687 & 0.625 \\
\midrule
15\% & 0.53 & 0.542 & 0.39 & 0.453 & 0.566 & 0.553 & 0.712 & 0.622 \\
\midrule
20\% & 0.518 & 0.521 & 0.439 & 0.476 & 0.534 & 0.523 & 0.837 & 0.644 \\
\midrule
25\% & 0.518 & 0.52 & 0.463 & 0.49 & 0.528 & 0.519 & 0.85	& 0.644 \\
\midrule
30\% & 0.518 & 0.52 & 0.475 & 0.496 & 0.534 & 0.522 & 0.862 & \textbf{0.65} \\
\midrule
35\% & 0.518 & 0.52 & 0.475 & 0.496 & 0.522 & 0.514 & 0.875 & 0.648 \\
\midrule
40\% & 0.524 & 0.526 & 0.487 & \textbf{0.506} & 0.522 & 0.514 & 0.875 & 0.648 \\
\bottomrule
\end{tabular}
}
\end{table}
For the evaluation, we generate API usage embeddings for a large set of historic code examples (from the pre-training data in Section~\ref{ss:pre-training-data}) for the MUBench APIs using both \murgcn{} and GraphCodeBERT and then cluster them, and map the MUBench examples to these clusters to predict correct usage or potential misuse based on cluster sizes.
We have used a Birch threshold of 3.0 based on our experiments and ablation studies. 
We experimented with multiple threshold values (e.g., $1.5$, $2.0$, $2.5$, $3.0$) and observed that increasing the threshold initially improves accuracy and precision, but values beyond 3.0 do not yield further gains. 
A threshold of 3.0 provided the best balance between precision, recall, and F1-score, which is why it was selected}. 
Specifically, while the $2.5$ threshold achieved an accuracy of $0.543$, precision of $0.574$, recall of $0.534$, and F1-score of $0.553$, the $3.0$ threshold improved performance to an accuracy of $0.584$, precision of $0.572$, recall of $0.687$, and an F1-score of $0.625$ (see the second row in Table~\ref{table:mubench-evaluation}). \textcolor{black}{We conducted a similar exercise for all baseline models and reported their best-performing versions.}

To distinguish between large and small clusters, we have experimented with multiple thresholds as shown in Table~\ref{table:mubench-evaluation}. 
The threshold is a percentage of the total number of examples for an API, which means if there are 100 examples that are clustered, and the threshold is $10$\% then the clusters having at least 10 examples will be considered as large (correct usage).


\textcolor{black}{As shown in Table~\ref{table:mubench-evaluation}, it reveals a clear trade-off between precision and recall as the threshold varies: lower thresholds (e.g., $10$\%) favour higher precision, while higher thresholds (e.g., $30$ - $35$\%) improve recall. 
A $30$\% threshold provides the best balance, yielding the highest F1-score for \murgcn{}.
\murgcn{} outperforms GraphCodeBERT across all metrics, achieving a best F1-score of $0.65$, which is $11$\% higher than GraphCodeBERT’s best F1-score of $0.506$.}


\begin{table}[t]
\caption{Performance of Misuse detectors on MUBench}
\label{table:other-detectors-on-mubench}
\centering
\footnotesize
\begin{tabular}{lrrr}
\toprule
\textbf{Detectors} & \textbf{Precision} & \textbf{Recall} & \textbf{F1-Score}\\
\midrule
MuDetect~\cite{MuDetect} & 33.00\% & 42.20\% & 36.96\% \\
\midrule
KGAMD~\cite{kgamd} & 60.00\% & 28.45\% & 38.59\% \\
\midrule
GraphiMuse~\cite{ma2024api} & 42.00\% & 54.50\% & 47.44\% \\
\midrule
Li et al.~\cite{Zhang-et-al} & \textbf{72.22\%} & 43.01\% & 53.91\% \\
\midrule
\murgcn{} (threshold = 30\%) & 52.2\% & \textbf{86.2}\% & \textbf{65.0\%} \\
\bottomrule
\end{tabular}

\end{table}

Table~\ref{table:other-detectors-on-mubench} presents the results reported by some of the popular misuse detectors.
\textcolor{black}{Using the F1-optimal operating point of $30$\%, \murgcn{} achieves $20.6\%$ improvement over the previous state-of-the-art by Li et al.~\cite{Zhang-et-al} on the MUBench dataset.}

\begin{table*}[t]
\caption{Performance of \mugnn{} with \& without sequence edges. The highest MI and RI scores are highlighted in each row.}
\label{table:without-se-abalation-study}
\centering
\small
\begin{tabular}{lrrrrrrrr}
\hline
\small
& \multicolumn{4}{c}{\textbf{Without Sequence Edges}} & \multicolumn{4}{c}{\textbf{With Sequence Edges}}\\
\cline{2-9}
\textbf{API methods} & \multicolumn{2}{c}{\textbf{\mugcn{}-Birch(1.5)}} & \multicolumn{2}{c}{\textbf{\murgcn{}-Birch(2.1)}} & \multicolumn{2}{c}{\textbf{\mugcn{}-Birch(1.5)}} & \multicolumn{2}{c}{\textbf{\murgcn{}-Birch(2.1)}}\\
\cline{2-9}
& \textbf{RI score $\uparrow$} & \textbf{MI score $\uparrow$} & \textbf{RI score $\uparrow$} & \textbf{MI score $\uparrow$} & \textbf{RI score $\uparrow$} & \textbf{MI score $\uparrow$} & \textbf{RI score $\uparrow$} & \textbf{MI score $\uparrow$}\\
\hline
Classloader.getResource() & 0.219 & 0.206 & 0.173 & 0.226 & 0.254 & 0.23 & \textbf{0.322} & \textbf{0.396}\\
\hline
Thread.start()& 0.114 & 0.231 & 0.124 & 0.143 &0.114&0.231&\textbf{0.439}&\textbf{0.472}\\
\hline
Statement.execute()& 0.065 & 0.165 & 0.211 & 0.268 &\textbf{0.34}&\textbf{0.422}&0.174&0.184\\
\hline
BufferedReader.read()& 0.056 & 0.117 & 0.248 & 0.326 &0.144&0.238&\textbf{0.412}&\textbf{0.492}\\
\hline
Timestamp.compareTo()& 0.39 & 0.489 & \textbf{0.492} & \textbf{0.492} &0.39&0.489&0.325&0.325\\
\hline
DataInputStream.readLine()&  0.445 & 0.5 & 0.289 & 0.315 &0.383&0.423&\textbf{0.41}&\textbf{0.465}\\
\hline
ServerSocket.bind()&  0.093 & 0.214 & \textbf{0.317} & \textbf{0.379} &0.098&0.198&0.231&0.298\\
\hline
ExecutorService.submit()& -0.016 & -0.032 & -0.014 & -0.016 &0.021&0.056& \textbf{0.278} & \textbf{0.306}\\
\hline
URI.getFragment() & 0.192 & 0.376 & 0.547 & 0.613 & 0.433 & 0.449 & \textbf{0.683} & \textbf{0.719} \\
\hline
Calendar.getTime() & 0.023 & 0.081 & \textbf{0.318} & 0.356 & 0.064 & 0.207 & 0.313 & \textbf{0.425} \\
\hline
Socket.connect() & 0.034 & 0.143 & 0.465 & 0.533 & 0.198 & 0.412 & \textbf{0.58} & \textbf{0.621} \\
\hline
JPanel.add() & 0.048 & 0.14 & 0.107 & 0.227 & 0.055 & 0.122 & \textbf{0.384} & \textbf{0.49}\\
\hline
FileChannel.read() & -0.015 & -0.019 & -0.014 & -0.014 & -0.007 & -0.007 & \textbf{0.142} & \textbf{0.157}\\
\hline
DateFormat.format() & 0.031 & 0.074 & 0.102 & 0.119 & 0.042 & 0.096 & \textbf{0.364} & \textbf{0.339}\\
\hline
ClassLoader.loadClass() & -0.041 & -0.072 & 0.057 & 0.068 & -0.041 & -0.072 & \textbf{0.208} & \textbf{0.232}\\
\hline
Runtime.freeMemory() & 0.009 & 0.026 & \textbf{0.213} & \textbf{0.282} & 0.007 & 0.059 & 0.151 & 0.175\\
\hline
Graphics2D.fill() & 0.006 & 0.026 & \textbf{0.188} & \textbf{0.229} & 0.013 & 0.071 & 0.004 & 0.02\\
\hline
DriverManager.getConnection() & \textbf{0.433} & \textbf{0.466} & 0.366 & 0.389 & 0.284 & 0.328 & 0.301 & 0.335\\
\hline
URL.openConnection() & 0.34 & 0.404 & 0.133 & 0.226 & 0.357 & 0.379 & \textbf{0.506} & \textbf{0.578}\\
\hline
File.toURI() & -0.022 & -0.056 & -0.014 & -0.015 & -0.02 & -0.038 & \textbf{0.171} & \textbf{0.171}\\
\hline
BufferedReader.readLine() & 0.095 & 0.139 & 0.118 & 0.224 & \textbf{0.216} & 0.284 & 0.183 & \textbf{0.304}\\
\hline
\textbf{Average} & 0.119 & 0.172 & 0.211 & 0.256 & 0.159 & 0.218 & \textbf{0.313} & \textbf{0.357}\\
\hline
\end{tabular}
\end{table*}

\subsection{RQ3: Impact of the Sequence Edges in AFG}
\label{ss:ablation-study-se}

A key novelty of our work is the AFG representation used by \mugnn{} to cluster API usage examples effectively.
AFG introduces sequence edges (SE) that capture API call order and read-read dependencies, such as enforcing that \texttt{file.close()} follows \texttt{file.open()}, which are not represented in standard data or control dependency graphs (e.g., those used in GraphCodeBERT).
We analyze the impact of SE edges on \mugnn{}'s performance in identifying API usage patterns.

Table~\ref{table:without-se-abalation-study} presents results with and without SE edges. 
%
Adding SE edges significantly improves the performance of \murgcn{}, yielding a 48\% and 40\% increase in RI and MI scores, respectively; \mugcn{} also shows notable gains.

Figure~\ref{fig:thread.start-example-1} and \ref{fig:thread.start-example-2} are two similar code examples using the API method {\tt Thread.start()}  where the thread is created, priority is set to minimum, and finally, the thread is started. \mugnn{} keeps these two examples in the same cluster when the SE edge between Lines 4 and 5 ({\tt Thread.setPriority()} to {\tt Thread.start()}) is there; otherwise, it considers the two usage examples different, illustrating the effect of SE edges.

\begin{figure}[t]
\begin{subfigure}{\linewidth}
\begin{lstlisting}[language=Java, numbers=left, numberstyle=\tiny, stepnumber=1, numbersep=3pt, xleftmargin=2em, breaklines=true, breakatwhitespace=true, basicstyle=\ttfamily\footnotesize]
public class Example{
  public void initDocumentCache(Book book){
    Thread documentIndexerThread = new Thread(new DocumentIndexer(book),                 "DocumentIndexer");
    documentIndexerThread. setPriority(Thread.MIN_PRIORITY); 
    documentIndexerThread.start();
  } }
\end{lstlisting}
\caption{{\tt Thread.start()} Example 1}
\label{fig:thread.start-example-1}
\end{subfigure}

\begin{subfigure}{\linewidth}
\begin{lstlisting}[language=Java, numbers=left, numberstyle=\tiny, stepnumber=1, numbersep=3pt, xleftmargin=2em, breaklines=true, breakatwhitespace=true, basicstyle=\ttfamily\footnotesize]
public class Example{
  public void BackgroundStreamSaver(InputStream in,OutputStream out){
    Thread myThread = new Thread(this, getClass().getName());
    myThread.setPriority(Thread.MIN_PRIORITY);
    myThread.start();
  } }

\end{lstlisting}
\caption{{\tt Thread.start()} Example 2}
\label{fig:thread.start-example-2}
\end{subfigure}
\caption{Two semantically similar usages of the {\tt Thread.start()} API method. 
}
\label{fig:thread-start-se-effect}
\Description{}
\end{figure}

\subsection{RQ4: Effect of AFG Pruning}
\label{ss:ablation-study-pruning}

Rather than using raw AFGs, we apply the pruning strategy described in Section~\ref{ss:afg-pruning} to remove nodes and edges unrelated to the API of interest. 
This step enhances API usage localization by retaining only relevant dependencies.
We evaluated \mugnn{}’s performance with and without pruning and found that, 
on an average, across all 21 APIs, \murgcn{} achieves an RI score of 0.313 and an MI score of 0.357  with pruning, compared to 0.298 and 0.346 without.
Thus, for \murgcn{}, pruning of AFGs improves the effectiveness, leading to a notable increase in both RI and MI scores.
We observed similar improvement for \mugcn{} also. 
The gains are smaller than those from SE edges, primarily because the API usage examples used for labelling and evaluation are typically small and focused on a single API, leaving fewer nodes and edges to prune.
\section{Discussion}
\label{s:discuss}

\noindent{\bf Threats to Validity.}
We evaluated clusters using APIs labeled by two expert programmers, who established labeling rules and resolved disagreements through discussion. 
However, a subsequent review by the authors, although finding most labels agreeable, identified areas for improvement, highlighting the challenge of fully eliminating human bias.
With additional resources (e.g., capital and manpower), the labeling process can be extended by involving more reviewers to supervise annotations, thereby reducing disagreements and improving the reliability of the ground-truth.

AFGNN, like any data-driven technique, reflects the usage patterns present in the code it uses. 
Therefore, when an API evolves and its usage protocol changes, the model may continue to capture outdated patterns until newer practices become prevalent in real-world code. 
This temporal dependency is a general limitation of data-driven and machine-learning-based approaches, rather than one specific to AFGNN.
In addition, our approach requires a sufficient number of usage examples for effective clustering.
For rarely used or unknown APIs, the lack of sufficient data can limit \mugnn{}’s applicability, a limitation shared by most LLM-based and static analysis approaches, which also rely on prior knowledge of the API's semantics and usage patterns. 
Thus, while \mugnn{} can generalize well to APIs with enough representative examples, its effectiveness may decrease for low-frequency or entirely novel APIs.

\noindent{\bf Design Decisions.}
We experimented with various small LMs, including CodeBERT, GraphCodeBERT, UnixCoder, and CodeT5+, as baselines. 
Among these, GraphCodeBERT and UnixCoder gave the best results, while CodeBERT and CodeT5+ produced significantly poorer clusters (with RI/MI scores close to 0 in most cases). 
GraphCodeBERT and UnixCoder performed well due to their consideration of token sequences, data flow, and AST information, emphasizing the importance of structural and semantic information for API usage tasks.
Although CodeT5+ underperformed in clustering, we use it for initializing node embeddings in \mugnn{} due to its strong performance in this context ($0.902$ accuracy, compared to $0.786$ of CodeBERT), with lower dimensionality ($256d$, compared to $768d$ of CodeBERT) and better computational efficiency. 

To determine the optimal number of clusters, we experimented with multiple internal validation metrics like the Davies-Bouldin Index (DBI)~\cite{davies-bouldin}, Silhouette score~\cite{Silhouettes}, and Calinski-Harabasz Index~\cite{CH_score}. 
Similarly, for external validation, we experimented with the Rand Index (RI)~\cite{rand-index}, Mutual Information (MI) score~\cite{mutual-info-score}, V-measure~\cite{v-measure}, and Fowlkes Mallows score~\cite{FM-score}. 
The correct metrics to use depends on the domain. 
Based on our experiments, we chose DBI for internal validation and RI, MI scores for external validation.

\section{Conclusion}

Understanding API usage patterns and detecting misuse can enhance security, saving time and effort.
Existing state-of-the-art models underperform in this task as they do not consider API-specific flow information.
\mugnn{} is effective by incorporating data flow, control flow, and API call sequences in code into a novel graph representation called the API Flow Graph (AFG).
It does not need any human intervention and can work with any API as it relies on vast open-source software repositories for usage examples.
Additionally, for any given API name, \mugnn{} can recommend frequent usage patterns by analyzing clusters of historical code examples, focusing on cluster centroids, and ranking these patterns based on cluster sizes.
This paper presents a detailed evaluation of \mugnn{}, demonstrating its superior performance in detecting potential misuse in real-world examples compared to state-of-the-art misuse detectors and {small} LMs. 



\bibliographystyle{ACM-Reference-Format}
\bibliography{main}


\end{document}